\documentstyle[12pt]{article}
\begin{document}

\def\lsim{\mathrel{\rlap{\lower3pt\hbox{\hskip0pt$\sim$}}
    \raise1pt\hbox{$<$}}}         
\def\gsim{\mathrel{\rlap{\lower4pt\hbox{\hskip1pt$\sim$}}
    \raise1pt\hbox{$>$}}}         

\renewcommand{\Im}{{\rm Im}\,}

\newcommand{\beq}{\begin{equation}}
\newcommand{\eeq}{\end{equation}}
\newcommand{\aver}[1]{\langle #1\rangle}

\newcommand{\La}{\overline{\Lambda}}
\newcommand{\Lam}{\Lambda_{\rm QCD}}

\newcommand{\lhs}{{\em lhs} }
\newcommand{\rhs}{{\em rhs} }

\newcommand{\ind}[1]{_{\begin{small}\mbox{#1}\end{small}}}
\newcommand{\hscale}{\mu\ind{hadr}}

\newcommand{\appa}{\mbox{\ae}}
\newcommand{\al}{\alpha}
\newcommand{\as}{\alpha_s}
\newcommand{\GeV}{\,\mbox{GeV}}
\newcommand{\MeV}{\,\mbox{MeV}}
\newcommand{\matel}[3]{\langle #1|#2|#3\rangle}
\newcommand{\state}[1]{|#1\rangle}
\newcommand{\ra}{\rightarrow}
\newcommand{\ve}[1]{\vec{\bf #1}}

\newcommand{\eq}[1]{eq.\hspace*{.15em}(\ref{#1})\hspace*{-.3em} }
\newcommand{\eqs}[1]{eqs.\ \hspace*{-.15em}(\ref{#1})}

\newcommand{\re}[1]{Ref.~\cite{#1}}
\newcommand{\res}[1]{Refs.~\cite{#1}}

\begin{titlepage}
\renewcommand{\thefootnote}{\fnsymbol{footnote}}

\begin{flushright}
UND-HEP-96-BIG\hspace*{.15em}08 \\
hep-ph/9612349\\
\end{flushright}
\vspace{1.9cm}
\begin{center} \LARGE
{\bf The Heavy Quark Expansion} 
\end{center}
\vspace*{.9cm}
\begin{center} 
{\Large
Nikolai  Uraltsev \footnote{Invited
talk at the Symposium on Radiative Corrections (CRAD\,96), Cracow, 
1--5 August \nolinebreak 1996.}}                          
\vspace*{.8cm}\\
{\normalsize 
{\it TH Division, CERN, CH 1211 Geneva 23, Switzerland},\\
{\it Dept. of Physics, Univ. of Notre Dame du Lac, Notre Dame, 
IN 46556, U.S.A.
}\\
and\\
{\it Petersburg Nuclear Physics Institute,
Gatchina, St.Petersburg 188350, Russia}
}\vfill

{\Large{\bf Abstract}}
\end{center}
\vspace*{.4cm}

I review the status of the modern theoretical approach to weak decays of heavy
flavor hadrons based on the $1/m_Q$ expansion in QCD.  The qualitative features
are explained and the subtleties in simultaneously  incorporating  perturbative
and power-suppressed effects are addressed. A few topical  phenomenological
applications are discussed in quantitative detail.

\vfill

\end{titlepage}
\addtocounter{footnote}{-1}

\newpage

\section{Introduction}

The Standard Model (SM) of fundamental interactions has 
so far yielded a consistent
description of various experimental data; still, from the general perspective
it is often viewed incomplete. A key role in exploring the SM is played by
studying electroweak decays of heavy flavor particles. Unique information
about the masses and mixing of quarks is obtained here thus setting the
stage for addressing some of the most mysterious questions about 
the origin
of the quark mixing and CP violation. 

The fact that the dynamics of heavy flavor hadrons 
represents a peculiar field where the strong interaction effects 
allow a systematic treatment based on first principles of QCD, 
was realized already in the 70s. Yet, the full variety of theoretical 
tools available for gauge theories was applied here only over the last few
years.

Two main directions were instrumental in developing QCD
towards its current status of the theory underlying strong interactions: one
was based on studying symmetry properties like isotopic or $SU(3)$ symmetry
and chiral invariance; the second exploited the asymptotic freedom which
allowed using dynamical calculations in the perturbative way for
high-energy processes. These two approaches, the `symmetry-based' and
`dynamical', are clearly seen in the evolution of the heavy quark theory.

During the early period to the end of the 80's, most applications 
carried the `dynamical' spirit and were done often at a somewhat
simplified intuitive level. The nonperturbative effects typically were 
thought to be small even in the decays of charm particles. At the same time,
the basic facts about the heavy quark spin and flavor symmetry were realized
and incorporated to the necessary extent. 

The symmetry considerations flourished for a few years with the formulation of
the so-called Heavy Quark Effective Theory (HQET) which made these symmetries
explicit at the level of the Lagrangian. It provided a convenient framework for
discussing exclusive semileptonic or electroweak radiative decays and
calculating basic perturbative corrections to them. On the other hand, the
range of applications of HQET was limited, and not all virtues of the
general Heavy Quark QCD Expansion (HQE) were employed. 

Finally, over the last few years a consistent well-defined dynamical
approach has been developed, which automatically respects the heavy quark
symmetries in a manifest way. It has been realized that the
nonperturbative corrections are often sizable even in the decays of beauty.
Incorporating the actual dynamical calculations allowed one to make the most
precise determinations of $|V_{cb}|$ and $|V_{ub}|$. 

The main effects in a weak decay of a heavy quark $Q$ originate 
from distances $\sim 1/m_Q \ll 1/\Lam$. Belonging to the 
weak coupling domain, $\as(m_Q)\ll 1$, they are tractable through 
perturbation theory. The QCD-interaction becomes strong only when the momentum
transfer is much smaller than the heavy quark mass, $k \ll m_Q$. This
simple observation elucidates the two basic ingredients of the heavy quark 
expansion that is genuinely based on QCD:
\begin{itemize}
\item The {\em nonrelativistic expansion}, which yields the effects of 
`soft' physics in the form of a power series in $1/m_Q$.
\item The treatment of the strong interaction domain 
based on the {\em Operator Product Expansion} (OPE).
\end{itemize}
Unless an analytic solution of  QCD is at hand, these two elements appear to be
indispensable for the heavy quark theory. 

The necessity of separating the two domains
characterized by momenta much larger than $\Lam$ and those which do not scale
with $m_Q$, is thus manifest. 
The perturbative corrections, on the other hand, bridge the
short-distance domain with the least understood domain of confining 
strong forces.

The general idea of separating two 
domains and applying different theoretical tools to them was formulated long
ago by K.~Wilson \cite{wilson} 
in the context of problems in statistical mechanics; in the
modern language, applied to QCD it is similar to lattice gauge theories. The
treatment of essentially Minkowski quantities has, however, some
peculiarities which attracted attention of theorists very recently.

The perturbative corrections -- in spite of being often complicated
technically -- look straightforward from the conceptual viewpoint. On the 
contrary,
getting even a minimal model-independent information about the nonperturbative
physics often looks like a miracle. In view of the nature of this
Symposium I will still try to put more accent onto the perturbative aspect, in
particular, since the perturbative corrections got a renewed 
attention during the
last two years. The overlap between the `perturbative' and `nonperturbative'
effects is non-trivial indeed.

There are different technical ways to perform $1/m_Q$ expansion in QCD; the
most popular schemes are known as the Nonrelativistic QCD (NRQCD) 
\cite{nrqcd,korner} and the Heavy Quark Effective Theory (HQET) \cite{hqet}.
In what
concerns the treatment of the perturbative corrections, in the standard HQET 
they are
simply added to what is considered as the nonperturbative physics. On the
other hand, the application of Wilson's idea requires separating of any
observable quantity into short-distance and long-distance pieces which,
formally, is a different procedure. 
Schematically, the HQET decomposition
$$
\mbox{An observable}\; = \; \mbox{perturbative piece} 
\;+\;\mbox{nonperturbative piece}
$$ 
is replaced in Wilson's OPE by a separation-scale--dependent
representation
$$
\mbox{An observable}\; = \; \mbox{short-distance piece}
\;+\;\mbox{long-distance piece}\;\;.
$$

The novel feature we face in the theoretical analysis of beauty decays
nowadays is that they often allow -- and even demand by virtue of the 
existence of
precise experimental measurements -- rather accurate predictions,
requiring a {\em simultaneous} treatment of perturbative and
nonperturbative QCD effects with the enough precision in both. This problem is
not new; the theoretical framework has been elaborated more than 10 years ago
\cite{fail}, but its phenomenological implementation was not mandatory 
until 
recently. It is the exceptional beauty of the fifth quark that allows -- in
spite of interference of the nonperturbative effects -- to make certain
reliable theoretical predictions with errors at a percent level, which
necessitates taking care of such `puristic' theoretical subtleties.
Nevertheless, failure to incorporate them properly, 
leads to certain theoretical
paradoxes and, unfortunately, some superfluous 
controversy in the numerical
estimates surfacing every now and then in the literature.

In spite of the significant progress made over the last years, the theory of
the heavy flavors is not yet a completed field and is undergoing to 
further extensive
development. There are quite a few topics which separately deserve a proper
review:
\begin{itemize}
\item Sum rules for the weak transitions with the nonperturbative effects
included. In a
certain way they replace the quantum mechanical (QM) description in the 
field theory. The perturbative 
corrections to the sum rules were actively discussed, often with quite
different conclusions.
\item The so-called BLM-resummation and renormalons.
\item The QCD-based OPE vs. `Fermi motion' in heavy quark decays.
\item The QCD sum rules calculations of the most interesting electroweak
formfactors in the $b\ra u$ ($b\ra s$) transitions.
\item Attempts to obtain 
constraints on the transition formfactors relying on the
most general unitarity and analyticity properties. Without an actual dynamical
input they are too weak, however. There is an interesting idea to
incorporate additional constraints from lattice QCD.
\item The heavy quark expansion for the production of heavy flavor hadrons
and quarkonia.
\end{itemize}

\noindent
The practical applications were in focus, too:
\begin{itemize}
\item Determination of $|V_{cb}|$ and $|V_{ub}|$.
\item A model-independent extraction of $m_b$.
\item Radiative corrections to ${\rm BR_{sl}}(B)$.
\end{itemize}
There is also a renewed interest in the inclusive nonleptonic widths.

Finally, the violations of duality is more and more discussed in the context of
the heavy quark physics.

Even this, by necessity truncated, list gives a feeling of the extensive
development taking place. 
Inevitably, I cannot cover here all, and even a large part of 
recent theoretical advances, or simply list all relevant references. A true 
measure of our understanding of the strong 
interaction dynamics is, eventually, determined by the theoretical accuracy
with 
which we can extract the fundamental underlying parameters like the KM mixing
angles from experimental data. Therefore, I will focus in this talk on a few 
selected
topics that illustrate the theoretical framework used in the most accurate
determination of the flavor mixing parameters, and review 
the overall status
of the heavy quark expansion for beauty particles, with the main emphasis on
the qualitative features. 

\section{Semileptonic decays}

The QCD-based heavy quark expansion can in principle equally be applied 
to all types of heavy flavor transitions. Semileptonic decays, 
however, are the
simplest case and I shall devote most of the attention to them. For practical
reasons I focus on the $b\ra c$ transitions; a brief discussion of the $b\ra u$
decays will be given later.

A typical semileptonic decay is schematically shown in Figs.~1. Generally, two
types of the decay rates can be 
singled out: the inclusive widths, where any combination
of hadrons is allowed in the final state, and the exclusive decays, when a
transition into a particular charmed hadron is considered, usually $D$ or
$D^*$.

\begin{figure}
\vspace{2.8cm}
\includegraphics{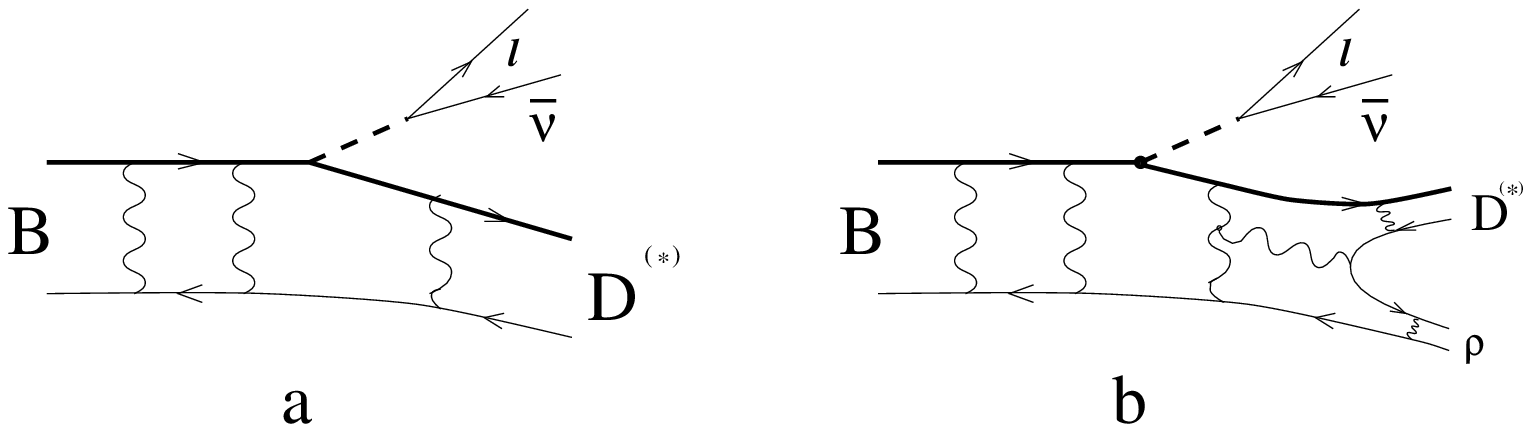}
\caption{
Exclusive $B\ra D^{(*)}$ ({\bf a})\hspace*{.5em} and generic
({\bf b})\hspace*{.5em} semileptonic decays.}
\end{figure}

\subsection{Inclusive semileptonic widths}

The four-fermion semileptonic decay width of a heavy quark has the form

\beq
\Gamma_{\rm sl} = \frac{G_F^2 m_b^5}{192\pi^3} |V_{cb}|^2 \cdot z_0
\left(\frac{m_c^2}{m_b^2}\right) \cdot \appa\;\;,
\label{1}
\eeq
where $z_0$ is the known phase space suppression factor and $\appa$ generically
includes all QCD corrections. In the heavy quark limit the difference resulting
from using the quark mass $m_b$ and the meson mass $M_B$ in \eq{1} disappears:
\beq
\frac{M_B-m_b}{m_b}\;\sim \; \frac{1}{m_b}\;\;.
\label{2}
\eeq
For the actual $b$ quark $m_b^5$ and $M_B^5$ differ by a
factor of $1.5$--$2$, which formally constitutes a power-suppressed 
nonperturbative
effect. This demonstrates the necessity of a systematic control of the 
nonperturbative corrections even in the decays of beauty particles.

The central result for the inclusive decay widths in QCD which is guaranteed 
by the application of the OPE to the heavy quarks in full QCD, 
is the absence of the $1/m_Q$ corrections \cite{buv,bs}. This seems
non-trivial {\em a priori} since there are such corrections to the 
masses, in particular, differentiating masses of different types of hadrons
($\Lambda_b$, $B$, $B^{**}$). The physical reason behind this fact is the
conservation of the color flow in QCD, that ensures the cancellation of the
effects of the color charge (Coulomb) interaction in the initial and final
states. In terms of 
nonrelativistic quantum mechanics it is the 
cancellation between the phase-space suppression caused by the Coulomb binding
energy in the initial state, and the Coulomb distortion of the final state
quark wavefunctions. The inclusive nature of the total widths leads to the fact
that the width is sensitive only to the strong 
interaction on the time scale $\sim
1/\Delta E \sim 1/m_b$ ($\Delta E$ denotes the energy
release). 
The final state interaction effect is thus not determined by 
the actual 
behavior of the strong forces at typical hadronic distances, but only by the
potential in the close vicinity of the heavy quark; the cancellation 
occurs universally, whether or not a nonrelativistic QM description is 
applicable.

The leading power corrections start with terms $1/m_b^2$; they have been
calculated in \re{buv,bs} and are expressed in terms of the expectation values
of two operators of dimension $5$, which in the language of QM are interpreted 
as 
the strength of the chromomagnetic field at the position of the heavy quark and
the square of the spacelike momentum of the heavy quark jiggling in the
rest-frame of the hadron, respectively:
\beq
\mu_G^2\simeq \frac{1}{2M_B}\matel{B}{\bar b \frac{i}{2}
\sigma_{\mu\nu}G^{\mu\nu} b}{B} \leftrightarrow \matel{B}{\vec{\sigma}_b
\cdot g_s\vec{{\cal H}_g}(0)}{B} 
\simeq \frac{3}{4} (M_{B^*}^2\!-\!M_B^2)\simeq 0.35
\GeV^2
\label{4}
\eeq
\beq
\mu_\pi^2\simeq \frac{1}{2M_B}\matel{B}{\bar b (i\vec{D}\,)^2b}{B} 
\leftrightarrow \matel{B}{\vec{p}^{\,2}}{B} 
\label{5}
\eeq
(the operators depend on the normalization point, but for
simplicity I do not indicate this fact). The size of $\mu_G^2$ for $B$ 
mesons is derived from the observed hyperfine mass 
splitting between $B$ and $B^*$.

The value of $\mu_\pi^2$ is not yet known directly; a model-independent 
lower bound was established in \cite{vcb,volkin}
\beq
\mu_\pi^2\; > \; \mu_G^2\;\;,
\label{6}
\eeq
which puts an essential constraint on its possible values. This bound is in 
agreement with QCD sum rule calculations \cite{pp},  
yielding a value about $0.5\GeV^2$ 
and with an estimate \cite{third} relying on the 
measured slope of the Isgur-Wise function.
In the absence of the perturbative gluon corrections, as 
it happens, for example, in 
simple QM models, the expectation value of the kinetic operator $\mu_\pi^2$
would coincide with the HQET parameter $-\lambda_1$; they are different,
however, in the actual field theory, where both $\mu_\pi^2$ and $\mu_G^2$
depend on the normalization point.

Including the nonperturbative corrections, the semileptonic width has the
following form \cite{buv,bs,bbsuv,prl}:
\beq
\Gamma_{\rm sl} = \frac{G_F^2 m_b^5}{192\pi^3} |V_{cb}|^2 \left\{z_0
\left(1-\frac{\mu_\pi^2-\mu_G^2}{2m_b^2} \right)
-2\left(1-\frac{m_c^2}{m_b^2} \right)^4\frac{\mu_G^2}{m_b^2}
-\frac{2}{3} \frac{\as}{\pi} z_0^{(1)} + ...
\right\}
\label{7}
\eeq
where ellipses stands for higher order perturbative and/or power corrections;
$z$'s are known phase space factors depending on $m_c^2/m_b^2$. Irrespectively
of the exact value of $\mu_\pi^2$, $\, 1/m_b^2$ corrections to $\Gamma_{\rm
sl}$ are rather small, about $-5\%$, thus leading to the increase in the
extracted value of $|V_{cb}|$ by $2.5\%$; the impact of the higher order power
corrections is negligible.

Good control of the QCD effects in the inclusive semileptonic widths provides
the most 
accurate direct way to determine $|V_{cb}|$ in a truly model-independent 
way. This method sometimes faces a traditional scepticism: which numerical 
value
must 
be used for $m_b$ and $m_c$? It appears that this practical problem has
deep roots; failure to understand them is the major source of
controversy about the heavy quark masses and inclusive widths often found 
in the literature. It
will be briefly discussed below.

In reality the precise value of $m_b$ is not too important, since the $b\ra c$
width depends to a large extent on the difference $m_b-m_c$ rather than on
$m_b$ itself; the former is constrained in the heavy quark expansion
\beq
m_b-m_c=\frac{M_B+3M_{B^*}}{4}-\frac{M_D+3M_{D^*}}{4} + \mu_\pi^2
\left(\frac{1}{2m_c}-\frac{1}{2m_b}\right) + ... \approx  3.50\GeV\;.
\label{8}
\eeq
It also independently enters lepton spectra in semileptonic decays \cite{prl}
and can be directly extracted from the data \cite{volspec}. Numerically
\cite{vcb,upset}, a change in $m_b$ by $50\MeV$ leads 
only to a $1\%$ shift in $|V_{cb}|$.\vspace*{.25cm}

\noindent
{\it Heavy quark masses}\vspace*{.15cm}

\noindent
An accurate knowledge of the heavy quark mass becomes important if a few
percent precision in $|V_{cb}|$ is targeted. However, rather different
estimates of $m_b$ can be found in the literature. The reason is that the most 
popular scheme where certain features of the HQE were implemented, the HQET 
was based on the
so-called pole mass of the heavy quarks. Not only was it a starting
parameter of the HQET-based expansions, it is this pole mass that one always 
attempted to extract from the experimental data. It
turns out, however, that the pole mass of the heavy quark is not a 
physical notion and cannot be even 
defined with the necessary accuracy in a motivated way: it suffers from 
an irreducible intrinsic theoretical 
uncertainty of order $\Lam$ \cite{gurman}. 
This applies therefore to the parameter 
$\La=M_B-m_b^{\rm pole}$ as well. 

At first sight this looks paradoxical and counter-intuitive: for the pole mass
is a usual mass of a particle; for example, the value of $m_{\rm e}$ quoted 
in the
tables of physical constants is just the {\em pole} mass of the electron. 
In QCD there is no `free heavy quark' particle in
the physical spectrum, and its pole mass is not well defined. 
The problems facing the possibilities to extract the pole mass from typical 
measurements were illustrated in \res{optical} and \cite{bsg}.

The physical origin of the uncertainty $\delta m_Q^{\rm pole} \sim \Lam$ is the
gluon Coulomb self-energy of the 
static colored particle. The energy stored in the
chromoelectric field inside a sphere of radius $R \gg 1/m_Q$ is given by 
\beq
\delta E_{\rm Coulomb} (R) \; \propto \;
\int_{ 1/m_b \sim |x| < R} \; \vec{E}_c^{\,2}
\;d^3x\;\; \propto \;\;{\rm const} - \frac{\as(R)}{\pi} \frac{1}{R} \;.
\label{9}
\eeq
The pole mass assumes that all energy associated with the color source is
counted, i.e. $R\ra \infty$. Since in QCD the interaction becomes strong at
$R_0 \sim 1/\Lam$, the domain outside $R_0$ would yield an uncontrollable (and
physically senseless) contribution to the mass $\sim \Lam$ \cite{gurman}.

Being a classical effect originating at a momentum scale well below $m_Q$, 
the appearance of this uncertainty can be traced in the usual perturbation
theory, where it manifests itself in higher orders 
as a so-called $1/m_Q$ infrared (IR) 
renormalon
singularity in the perturbative series for the pole mass \cite{pole,bbpole}.

\begin{figure}
\vspace{2.8cm}
\includegraphics{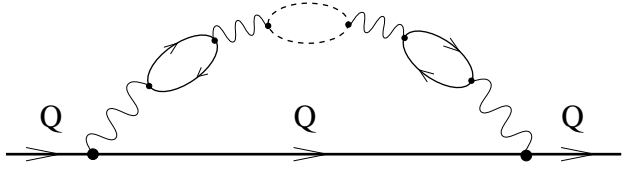}
\caption{
Perturbative diagrams leading to the IR renormalon
uncertainty in $m_Q^{\rm pole}$ of the order of $\Lam$. The contribution of
the gluon momenta below $m_Q$ expresses the classical Coulomb self-energy of
the colored particle. The number of bubble insertions into the gluon propagator
can be arbitrary.}
\end{figure}

To model the effect of the running coupling and emergence of the
nonperturbative low-momentum domain one can use the running coupling
$\alpha_s(k^2)$ in the one-loop diagram for the correction to the on-shell
mass of the heavy quark, see Fig.~2. 
In the nonrelativistic limit $k\ll m_Q$ the expression
simplifies and reads
\beq
\delta m_Q \; \sim \; -\frac{4}{3}\, \int \;\frac{d^4 k}{(2\pi)^4 i}\:
\frac{4\pi\as(k^2)}{k^2} \,\frac{1}{k_0}\;=\;
\frac{4}{3}\, \int \;\frac{d^3 \vec{k}}{4\pi^2 }\:
\frac{\as(\vec{k}^{\,2})}{\vec{k}^{\,2}} \;.
\label{9a}
\eeq
Expanding running $\as(k^2)$ in terms of a short-distance $\as(\mu^2)$ one
obtains for $\delta m_Q$ the whole series in $\as(\mu^2)$. The coefficients
grow factorially:
\beq
\frac{\delta m_Q}{m_Q} \; \sim \; \frac{4}{3}\, \frac{\as}{\pi}\, 
\left(\frac{b\as}{2\pi} \right)^n\, n!\;\;,\qquad \qquad\qquad 
b=\frac{11}{3} N_c-\frac{2}{3} n_f = 9\;,
\label{9b}
\eeq
and the nonperturbative regime manifests itself as the 
uncertainty in defining a sum of such a series. 
Numerically, the calculation of the contribution of the domain below $\mu\simeq
1\GeV$ would then look as follows:
$$
m_b^{\rm pole} \; =\; m_b(1\GeV) +\delta m^{\rm pert} \; \simeq \;
$$
\beq
4.55\GeV+0.25\GeV + 0.22\GeV + 0.38\GeV
+ 1\GeV + 3.3\GeV + 14\GeV +\:...
\label{9c}
\eeq
Clearly, such a possibility does not look encouraging if one really needs to
know the heavy quark mass accurately enough!

The first two terms in \eq{9c} constituting approximately $4.8\GeV$ can be
called -- with some reservations \cite{optical} -- the `one-loop' pole mass
that enters typical calculations when order-$\as$ corrections are computed.
Recently, the third term was incorporated \cite{gremmlig} in the analysis of the
lepton spectra in the semileptonic $b$-decays and the value
\beq
\La \;=\; M_B-m_b^{\rm pole} \; \simeq \; (390\pm 110)\MeV
\label{9d}
\eeq
was deduced. The quoted error, according to \cite{gremmlig}, includes ``only
experimental uncertainty'' -- however, such a numerical conclusion 
can hardly be taken sensibly.

In spite of this irreducible uncertainty in $m_Q^{\rm pole}$ and $\La$, 
the inclusive widths, though dependent on the mass, can be
theoretically calculated since they are governed not by the pole masses, 
but well-defined short-distance running masses $m_Q(\mu)$ with the  
Coulomb energy originating from distances $\gsim 1/\mu$ peeled off. Moreover, 
it is precisely 
this short-distance running mass rather than the pole mass that can be
extracted from experiment with, in principle, unlimited accuracy: the pole mass
does not enter any genuine short-distance observable at the level of
nonperturbative corrections, which can be calculated by
means of the OPE \cite{pole}. 
It is worth noting that the irrelevance of the pole mass roots deeper than
merely the problem of the IR renormalon; even if there were no problem in
defining the pole mass in QCD, its infrared part distinguishing it from the
short-distance mass is still foreign to the OPE.

Applied to the inclusive widths, this observation suggests certain information
about the importance of higher order perturbative corrections: if masses
entering \eq{7} are the pole masses, the perturbative series 
\beq
\Gamma_{\rm sl}^{\rm pert} = \Gamma_0 \appa^{\rm pert}=\Gamma_0\left(1+
a_1 \frac{\as}{\pi} + a_2 \left(\frac{\as}{\pi}\right)^2 +...
\right)
\label{10}
\eeq
is poorly behaved,
with coefficients $a_k$ factorially growing, which makes it non-convergent and
the radiative correction factor {\em per se} uncalculable in principle either 
with an 
accuracy $\sim \Lam/m_b$. In contrast, if one uses the
short-distance masses, the higher-order corrections become smaller and the
overall factor is theoretically calculable with the necessary precision
\cite{pole,bbz}.

This, at first glance rather academic, observation in reality proved to
underlie the pattern of the corrections from the very 
first terms. Remarkably, the actual model-independent calculations 
of $\Gamma_{\rm sl}$ through observables
measured in experiment are very stable against perturbative corrections.
Including ${\cal O}(\as^2)$ terms in the extraction of the $b$ pole mass from,
say, the $\rm e^+e^- \ra \bar b b$ threshold region \cite{volmb} noticeably
increases its value. However, if one performs the parallel perturbative
improvement in calculating the width, one finds an essential suppression of the
perturbative factor $\appa^{\rm pert}$. The two effects offset each other 
almost completely.

This conspiracy is not unexpected: the appearance of large corrections at both
stages is merely an artefact of using the ill-defined pole mass in the
intermediate calculations. 
The situation is peculiar since the actual
nonperturbative effects appear only at the level $1/m_Q^2$, whereas the pole
mass is infrared ill-defined already at an accuracy of $1/m_Q$.
The failure to realize this fact led to a superficial
suggestion \cite{savage} that even in beauty particles the perturbative
corrections may go out of theoretical control; a more careful analysis
\cite{upset,bbbsl} showed that this is not the case.

Since the apparent troubles with the perturbative corrections are associated 
with 
the pole masses, it is better to get rid of them altogether using 
instead running
masses $m_{b,c}(\mu)$ normalized at $\mu \sim 1\GeV$; this is advantageous both
theoretically and in practice. This has been done in \cite{upset} and
showed that neither masses nor the perturbative corrections to the width
$\appa^{\rm pert}$ have 
significant contributions from higher orders. It is important to
note that it is not a kind of empirically selected procedure but is required in
the literal implementation of Wilson's OPE in the case of the heavy quark
widths \cite{pole}.

To summarize, the idea that the perturbative corrections in the
extraction of $|V_{cb}|$ from $\Gamma_{\rm sl}(B)$ are large emerges from an
attempt to use ill-defined pole masses in an inconsistent way where two
problems are encountered:
\begin{itemize}
\item It is 
`difficult to extract' accurately $m_b^{\rm pole}$ from 
experiment; in any
particular calculation the effects are easily identified that were left out, 
which can
change its value by $\sim 200\MeV$. This uncertainty leads to a `theoretical
error' $\delta_{\rm I}$ in $\Gamma_{\rm sl}(B)$ of about $10\%$.
\item When routinely calculating $\Gamma_{\rm sl}(B)$ in terms of
the 
pole masses, there are significant higher order corrections $\delta_{\rm II} 
\simeq 10\%$. 
\end{itemize}

The naive conclusion drawn from such experience \cite{neubmori} is
that one cannot reliably calculate the width without $\sim 20\%$ uncertainty:
$$
\frac{\delta 
\Gamma_{\rm sl}}{\Gamma_{\rm sl}}= \delta_{\rm I} + \delta_{\rm II} 
\simeq 20\%\;
\; \leftrightarrow \;\;\frac{\delta |V_{cb}|}{|V_{cb}|} \simeq 10\%\;\;.
$$
On the contrary, theory predicts a cancellation 
between $\delta_I$ and
$\delta_{II}$ in a consistent perturbative 
calculation, and that was explicitly checked in
\cite{upset,bbbsl}. For example, the net impact of the calculated (presumably
dominant) second-order ${\cal O}(\as^2)$ corrections on the value of $|V_{cb}|$
appeared to be less than $1\%\,$! Moreover, just neglecting {\em all}
perturbative corrections altogether, both in the semileptonic width and
in extracting $m_b$ from experiment, yields a $|V_{cb}|$ smaller by less than 
$5\%$ \cite{volpriv}. \vspace*{.2cm}

Recently, all-order corrections
associated with the effect of 
running of $\as$ in one-loop diagrams (I shall
refer to it in what follows as an extended BLM, or, shorter, merely a BLM 
approximation) were calculated in \cite{bbbsl}, thus improving the exact
one-loop result which has been known from the QED calculations \cite{muon}.
Using the most accurate model-independent low-energy determination of $m_b$ 
\cite{volmb} one gets \cite{upset}
$$
|V_{cb}|=0.0413\left(\frac{{\rm BR}(B\rightarrow X_c\ell\nu)}{0.105}
\right)^{\frac{1}{2}}\left(\frac{1.6\,\rm ps}{\tau_B}\right)^{\frac{1}{2}}
\times
$$
\beq
\left(1-0.012\frac{(\tilde \mu_\pi^2-0.4\,\rm GeV^2)}{0.1\,\rm
GeV^2}\right)\cdot \left(1-0.006\frac{\delta m_b^*}{30\,\rm MeV}\right)\;.
\label{26b}
\end{equation}
The relevant sources of theoretical uncertainty are shown explicitly. The
main one is lack of knowledge of the exact value of $\mu_\pi^2$ (marked with a 
tilde 
in \eq{26b}, indicating that a particular field-theoretic definition of the
kinetic operator has been assumed), which enters
through the value of $m_b-m_c$, eq.~(\ref{8}). 
A dedicated analysis of the lepton spectra will
reduce this uncertainty. At the moment a reasonable estimate of the 
actual uncertainty in $\mu_\pi^2$ is about $0.2\GeV^2$, leading to a 
$2.5\%$ uncertainty in $|V_{cb}|$.

The dependence on the absolute value of $m_b$ is minor; since we rely here on 
the
well-defined short-distance mass $m_b^*$, there is no intrinsic
uncertainty in its value. 
The analysis \cite{volmb} estimated $\delta m_b^*
\simeq 30\MeV$ which seems reasonable; to be confident, I double it, and even
then the related uncertainty in $|V_{cb}|$ is less than $1.5\%$.

Finally, we should consider the perturbative corrections. As 
previously explained, 
the actual impact of the known perturbative corrections when 
relating the
semileptonic width to other low-energy {\em observables} 
is very moderate, and there
is no reason to expect the higher-order effects to be significant. With the 
{\em a priori} dominant all-order BLM corrections calculated \cite{bbbsl}, one
may 
be concerned only with the true two-loop effects ${\cal O}(\as^2)$ which
are not associated with the running of $\as$. These have not been calculated
completely yet; however, the recent calculation \cite{czar} of the 
complete ${\cal O}(\as^2)$ corrections in the small velocity kinematics 
demonstrated the feasibility of a complete 
calculation and suggested that the corrections must be 
small. There are some enhanced higher-order corrections not
related to running of $\as$, that are specific to 
the inclusive widths
\cite{ive}. They have been accounted for in the analyses \cite{vcb,upset}, but
went beyond those in \cite{bn,bbbsl}. With the results of \cite{czar} it seems
unlikely that as yet uncalculated second-order corrections can change the 
width by
more than $2$--$3\%$; therefore, assigning an additional uncertainty of $2\%$ in
$|V_{cb}|$ is a quite conservative estimate.

Adding up these uncertainties we conclude that at present we can confidently
calculate $|V_{cb}|$ from $\Gamma_{\rm sl}$ with the theoretical uncertainty
\beq
\frac{\delta|V_{cb}|}{|V_{cb}|}\vert_{\rm th}\; <  \;5\%\;\;.
\label{17}
\eeq
Keeping in mind some controversy about experimental value of ${\rm BR_{sl}}(B)$
and certain changes in time in $\tau_B$, it is fair to say that the
theoretical accuracy in extracting $|V_{cb}|$ is even better than its current
experimental counterpart.

It is important that this method has a potential for further improvement in a 
model-independent way. 
I think that the $2\%$ level
of a {\em defensible} theoretical precision can be ultimately reached here.
It is difficult to 
count on an essential
improvement beyond that, because of impact of higher-order power 
corrections and
possible violations of duality.

In a similar way it is straightforward 
to relate the value of $|V_{ub}|$ to 
the total semileptonic width $\Gamma(B\ra X_u\, \ell\nu)$ \cite{upset}:
\beq
|V_{ub}|=0.00458\left(\frac{{\rm BR}(B\rightarrow X_u\ell\nu)}{0.002}
\right)^{\frac{1}{2}}\left(\frac{1.6\,\rm ps}{\tau_B}\right)^{\frac{1}{2}}\;.
\label{18}
\eeq
An accurate measurement of the 
inclusive $b\ra u\, \ell\nu$ width is difficult and
for a long time seemed unfeasible. However, recently ALEPH announced the first
direct measurement \cite{aleph}: 
$$
{\rm BR}(B\rightarrow X_u\ell\nu) = 0.0016\pm 0.0004\;\;.
$$
I cannot judge the reliability of the quoted error bars in this 
complicated analysis; it certainly will be clarified in the future.
Accepting the numbers literally, I arrive at the model-independent result
\beq
\frac{|V_{ub}|}{|V_{cb}|}\; = \; 0.098\pm 0.013\;\;.
\label{19}
\eeq
The theoretical uncertainty in translating $\Gamma(B\ra X_u\, \ell\nu)$ into
$|V_{ub}|$ is a few times smaller and is not seen in the final result.

To conclude this section on the inclusive semileptonic widths, let me briefly
comment on the literature. It is sometimes stated that the existing uncertainty
in $\Gamma_{\rm sl}$ is at least $20\%$ \cite{neubmori,neubert}. The 
major origin of such
claims is ignoring the subtleties related to 
using the pole mass in the
calculations, and considering separately the perturbative corrections to the
pole masses and to the widths expressed in terms of $m_Q^{\rm pole}$. This is
inconsistent on theoretical grounds \cite{pole}, whose relevance was 
confirmed by the
concrete numerical evaluations of the BLM corrections in 
\cite{upset,bbbsl}. The
dependence on $m_b$ and $m_b-m_c$ used to determine the uncertainty in
$|V_{cb}|$ was calculated erroneously in \cite{neubmori} (cf. \cite{upset}), 
apparently because of an arithmetic mistake that led to a significant 
overestimate.
Finally, no convincing argument was 
given to justify a sevenfold boosting of
the theoretical uncertainty in $m_b$ obtained in the dedicated analysis
\cite{volmb}.

\subsection{Exclusive zero-recoil $B\ra D^*\,\ell \nu$ rate}

Good theoretical 
control of all QCD effects in the inclusive widths was due
to the fact that removing constraints on the final state to which
decay partons can hadronize, makes such a probability a truly short-distance
quantity amenable to a direct OPE expansion. Application of a 
similar idea to the exclusive zero-recoil decay rate $B\ra D^*\,\ell \nu$
yielded quite an accurate determination of $|V_{cb}|$ as well 
\cite{vcb,optical},
though with a more significant irreducible model dependence and a larger
intrinsic uncertainty. The limitation is twofold: first, constraining the
decays to a specific final state makes the transition not a genuinely
short-distance effect; second, it suffers from 
downgrading the 
expansion parameter, namely $1/m_c$ rather than $1/m_b$. 

On the quark level the decay is shown in Fig.~1a; near zero recoil it is
determined by the single hadronic formfactor $F_{D^*}$. In the absence of
corrections violating the heavy quark symmetry, $F_{D^*}=1$ holds; 
for finite $m_{b,c}$ it acquires corrections:
\beq
F_{D^*} = 1 -(1-\eta_A)+ \delta_{1/m^2} + ...\;\;.
\label{22}
\eeq
The effect of the nonperturbative domain denoted by the last term and the
ellipses starts with the terms $\sim 1/(m_c,m_b)^2$ \cite{vs,luke}, but
otherwise is rather arbitrary, since it depends on the details of the 
long-distance
hadronization dynamics in the form of wavefunction overlap. This opened 
the field
for speculations and certain controversy about the value of $F_{D^*}$
\cite{neubpr}.

The situation as it existed by 1994 was summarized in the review lectures
\cite{neubtasi}:
\beq
\eta_A=0.986\pm 0.006 \qquad \qquad \qquad
\delta_{1/m^2} =(-2\pm 1)\% \qquad ,
\label{23}
\eeq
yielding $F_{D^*}\simeq 0.97$, and was assigned the status of ``one of the most
important and, certainly, most precise predictions of HQET". Nowadays we
believe that the actual corrections to the symmetry limit are larger, and the
central theoretical value lies rather closer to $0.9$ \cite{vcb,optical}.

Regarding the perturbative calculations {\em per se}, it was later pointed out 
\cite{comment} that the improvement \cite{neubimp} of
the original one-loop estimate was incorrect, and the proper value is
rather $\eta_A\approx 0.965 \pm 0.025$; subsequent calculations of the 
higher-order 
corrections in the BLM approximation \cite{bbbsl,flaw}  confirmed it: 
$\eta_A\approx0.965 \pm 0.02$. The purely perturbative chapter was 
closed recently with the complete two-loop ${\cal O}(\as^2)$ result
\cite{czar}
\beq
\eta_A^{(\rm 2\, loop)} =0.960 \pm 0.007\;\;.
\label{24}
\eeq
It should be noted, however, that the inherent irreducible uncertainty of the
{\em complete} perturbative series for $\eta_A$ 
exceeds the one stated in \eq{24} by a factor of three \cite{bbbsl,ns,pert}.

If the mass of the charm quark were a few times larger, from the practical
viewpoint the two-loop calculation would have been the whole story for
$F_{D^*}$. In reality, the power corrections originating from the domain of
momenta below $\sim 0.6\GeV$ appear to be more significant. Unfortunately, 
not much can be said about them 
in a model-independent way, although they have been shown to be negative and
exceed about $0.04$ \cite{vcb,optical} in magnitude (the analysis for
$\Lambda_b$ was given in \cite{flambda}).

The idea of this dynamical 
approach was to consider the sum over all possible hadronic
states in the zero-recoil kinematics but not to limit it to only $B\ra D^*$;
such a rate sets an upper bound for 
the production of $D^*$. This {\em
inclusive} quantity is of a short distance nature and can be calculated in QCD
by means of the OPE. Schematically, the result through order $1/m^2$ is
\beq
|F_{D^*}|^2 +\sum_{\epsilon_i<\mu} |F_i|^2 = \xi_A(\mu)
 -\frac{\mu_G^2}{3m_c^2} -
\frac{\mu_\pi^2-\mu_G^2}{4}
\left(\frac{1}{m_c^2}+\frac{1}{m_b^2}+\frac{2}{3m_cm_b}
\right)\;\;,
\label{25}
\eeq
where $F_i$ are the-axial current transition formfactors to excited charm
states $i$ with the mass $M_i=M_{D^*}+\epsilon_i$, and $\xi_A$ is a
perturbative renormalization factor (the role of $\mu$ will be 
addressed later). Considering
a similar relation for another type of `weak current', say, $\bar{c}
i\gamma_5 b$, one obtains a different sum rule 
\beq
\sum_{\tilde \epsilon_k<\mu} |\tilde F_k|^2 = 
\left(\frac{1}{m_c}-\frac{1}{m_b}\right)^2
\frac{\mu_\pi^2-\mu_G^2}{4}\;\;,
\label{26}
\eeq
with the tilde 
referring to the quantities for the transitions induced by 
this hypothetical current. These sum rules (and similar ones at arbitrary
momentum transfer), established in \res{vcb,optical}, have been subjected to
a critical scrutiny for two years, but are now accepted and constitute 
the basis for currently used estimates of $F_{D^*}$.

Since \eq{26} is the sum of certain transition probabilities, it is
definite-positive and results in a rigorous lower bound
\beq
\mu_\pi^2\;>\;\mu_G^2 \simeq 0.4\GeV^2\;\;.
\label{27}
\eeq
The sum rule (\ref{25}) then leads to the model-independent lower bound for the
$1/m^2$ terms in $F_{D^*}$:
\beq
-\delta_{1/m^2}\;>\;\frac{M_{B^*}^2-M_B^2}{8m_c^2}\simeq 0.035\;.
\label{28}
\eeq
The actual estimate depends essentially on the value of $\mu_\pi^2$. It was
suggested in \cite{vcb} to estimate the contribution of the excited states in
the l.h.s. of the sum rule (\ref{25}) from $0$ to $100\%$  of the power
corrections in the  r.h.s.:
\beq
-\delta_{1/m^2}=(1+\chi)\left(\frac{M_{B^*}^2-M_B^2}{8m_c^2}
+\frac{\mu_\pi^2-\mu_G^2}{8}
\left(\frac{1}{m_c^2}+\frac{1}{m_b^2}+\frac{2}{3m_cm_b} \right)
\right)\;,\;\; 0\le \chi \le 1\;.
\label{29}
\eeq
If so, one arrives at \cite{vcb}
$$
-\delta_{1/m^2}= (5.5\pm 1.8)\% \qquad \mbox{ at } \; \mu_\pi^2=0.4\GeV^2
$$
\beq
\,-\delta_{1/m^2}= (6.8\pm 2.3)\% \qquad \mbox{ at } 
\mu_\pi^2=0.5\GeV^2\;\mbox{ }
\label{30}
\eeq
$$
-\delta_{1/m^2}= (8.1\pm 2.7)\% \qquad \mbox{ at } \; \mu_\pi^2=0.6\GeV^2
$$

Before proceeding to further phenomenological implications, let me dwell 
on the QM meaning of the sum rules \cite{optical}. The
act of a weak semileptonic 
decay of the $b$ quark is its instantaneous replacement by $c$
quark. In ordinary QM the overall probability of the produced state to
hadronize to some final state is exactly unity, which (neglecting radiative
corrections) is the first, leading term in the r.h.s. of (\ref{25}). 
Why then are there  
nonperturbative corrections in the sum rule? The answer is that
the normalization of the weak current $\bar c \gamma_\mu \gamma_5 b$ is not
exactly unity and depends, in particular, on the external gluon field.
Expressing the QCD current in terms of the nonrelativistic fields used in 
QM one has \cite{optical,korner}, for example, 
\beq
\bar c \gamma_k \gamma_5 b \leftrightarrow \sigma_k -
\left(\frac{1}{8m_c^2}(\vec\sigma
i\vec D)^2\sigma_k+\frac{1}{8m_b^2}\sigma_k(\vec\sigma
i\vec D)^2\;-
\frac{1}{4m_cm_b}(\vec\sigma i\vec D )\sigma_k(\vec\sigma i\vec D) 
\right) +
{\cal O}\left(\frac{1}{m^3}\right)
\label{32}
\eeq 
The last term just yields the correction seen in the \rhs of the
sum rule.

The inequality $\mu_\pi^2 >\mu_G^2$ in QM expresses the positivity of the Pauli
Hamiltonian 
$$
\frac{1}{2m}(\vec\sigma\,i\vec D\,)^2 =
\frac{1}{2m}\left((i\vec D)^{\,2}-\frac{i}{2}\sigma G\right)
$$
\cite{volkin}.  It 
is interpreted as 
the Landau precession of a charged (colored) particle in the
(chromo)magnetic field where one has $\aver{p^2}\ge |\vec B|$. 
Literally, in $B^*$ the QM expectation value of the chromomagnetic field is 
suppressed, only
$\mu_G^2/3$, and in $B$ it vanishes, since 
$\vec{B}$ is proportional to the spin of the light degrees of freedom. However, 
essentially non-classical nature of $\vec{B}$ (e.g., 
$\aver{\vec{B}^{\,2}} \ge 3 \aver{\vec{B}}^{2}$) in turn 
enhances the bound  which then takes the same form as  in the external
classical field.

Before returning to numbers, let me emphasize that the perturbative factor
$\xi_A$ is not (and cannot be) equal to $\eta_A^2$ \cite{optical} (see also
\cite{pole}). It is clear that purely perturbative calculations include
(although in an improper way) 
the effect of the strong interaction domain completely 
described by the power terms. $\xi_A$ depends explicitly 
on the separation scale
$\mu$ dividing 
the two domains. Similarly, the values of $\mu_\pi^2$ and
$\mu_G^2$ depend, in fact, on $\mu$ (which, for simplicity, 
was neglected above).

Most important, unlike $\eta_A$ which in principle cannot be defined
theoretically with better than a few percent accuracy, $\xi_A(\mu)$ is a 
well-defined quantity and can be calculated in the small-coupling expansion
provided $\mu$ is large enough. It is fair to say that no
significant uncertainty can be associated with this factor which is close to
unity, $\xi_A \simeq (0.99)^2$.

What can be concluded for $F_{D^*}$ numerically? Even allowing a very moderate
interval for the yet unknown value of $\mu_\pi^2$ between $0.4\GeV^2$ and
$0.6\GeV^2$, we see $-\delta_{1/m^2}$ varying between $3.5\%$ and $11\%$;
moreover, since there are hardly any model-independent arguments to prefer any
part of the interval {\em a priori}, the whole range must be considered equally
possible. Adding small perturbative corrections we end up with the 
reasonable estimate $F_{D^*}\approx 0.9\,$.
It is curious to note that at a `central' value $\chi=0.5$
the dependence of the zero-recoil decay rate on $\mu_\pi^2$ 
through $\delta_{1/m^2}$ 
practically coincides with that of $\Gamma_{\rm sl}(B)$, see \eq{26b}, although
they actually change in the opposite directions.
 
The typical size of the $1/m_Q^2$ corrections to the exclusive zero-recoil
decay rate seems to be significant, around $15\%$, which 
is expected since they are driven by the 
scale $m_c\simeq 1.3\GeV$. It is evident that $1/m_c^3$
correction in $F_{D^*}$ not addressed so far are {\em at least} about
$\frac{1}{2}(0.15)^{3/2} \simeq 2\!-\!3\%$.\footnote{This is consistent 
with the fact that the $1/m_Q^3$ IR
renormalon ambiguity in $\eta_A^2$ constitutes $5\%$ at $\Lambda_{\rm QCD}^{\rm
\overline{MS}} \simeq
220\MeV$ \cite{pert}.}

Thus, I believe that the current theoretical technologies do not allow to
reliably predict the zero-recoil formfactor $F_{D^*}$ with a precision better 
than
$5$--$7\%$ in a model-independent way; its value is expected to be
approximately $0.9$, although a correction to the symmetry limit twice 
smaller, 
as well as larger deviations, are possible. It is encouraging that the `educated
guess' $F_{D^*}\simeq 0.9$ which emerged from the first -- and 
so far the only -- 
dynamical QCD-based consideration
\cite{vcb,optical}, yielded a value of $|V_{cb}|$ rather close
to a less uncertain result obtained from $\Gamma_{\rm sl}(B)$, although the
experimental error bars for the zero-recoil rate are still significant and its
status does not seem finally settled yet.

In future, more accurate data will enable us to measure $F_{D^*}$
with a theoretically informative precision using $|V_{cb}|$ from $\Gamma_{\rm
sl}(B)$, and thus provide us with deeper insights into the dynamics of strong
forces in the heavy quark system. 

Smaller theoretical uncertainty,
$2.5\%$ and $3\%$, 
is now quoted by Neubert for $\delta_{1/m^2}$ and $F_{D^*}$,
respectively. The
former was obtained in
\cite{update},  in what he calls a ``hybrid approach'',\footnote{I disagree 
with
the statements of \cite{update},
reiterated in later papers, suggesting that the original analysis
\cite{vcb,optical} missed some elements of the heavy quark spin-flavor
symmetry;
on the contrary, it was stated in the latter paper that all these relations
automatically emerge from the sum rules 
that replace the QM wavefunction description in the quantum field theory.} 
which reduces to
assigning the fixed value $\mu_\pi^2=0.4\GeV^2$ and using
it in the sum rule (\ref{25}) within the same model assumption 
$0\le \chi\le 1$ suggested in \cite{vcb}.
Correspondingly, the quoted number
for $\delta_{1/m^2}$ practically
coincided with the first line of \eqs{30}. 
In reality, allowing
$\mu_\pi^2$ to vary within any reasonable interval
significantly stretches the uncertainty.
Moreover, the analysis \cite{update} was based on using $\eta_A$ as the 
perturbative factor assuming, literally, that in the proper treatment the final
result would not be changed numerically -- which is just the case according to
consideration of \re{lig}. On top of that, the uncertainty in the definition of
$\eta_A$ due to $1/m^2$ and $1/m^3$ IR renormalons constitutes $2$--$3\%$ 
each and
thus is an additional one in the usage of the sum rules adopted in
\cite{update}. As a result, the stated theoretical
accuracy of those estimates cannot be accepted as realistic.

\section{Perturbative corrections to the sum rules and the IR renormalons}

The perturbative corrections have been mentioned already in the context of the
semileptonic widths; the nonperturbative effects there were small. A
different situation occurs when the latter are in the focus; the interplay
between the perturbative and nonperturbative physics becomes important. This
happens in the analysis of the sum rules, in particular, when the effects of
the running of $\as$ is included.

A typical short-distance observable $A$ at the one-loop level can be
schematically written in the form
\beq
A(Q^2)\; \sim \; 1\;+\; \int \;\frac{d^4 k}{4\pi^3 }\:
\phi\left(\frac{k^2}{Q^2}\right)\, \as(k^2)
\label{c4}
\eeq
where $\phi$ is some function and 
$Q^2$ sets the high momentum scale. Even at $Q^2 \gg \Lam^2$, small 
momenta $k^2 \sim \Lam^2 \simeq Q^2\, {\rm e}\,^{-4\pi/(b\as)}$ contribute at a 
certain level, but in this
nonperturbative domain the coupling $\as(k^2)$ cannot be expanded in terms of
$\as(Q^2)$. This leads to emergence of the factorial growth of the
coefficients in the perturbative expansion of $A$:
\beq
A(Q^2)\; \sim \; 1\;+\; \sum_{n=1}^{\infty} \;a_n\,
\left(\frac{\as}{\pi}\right)^n\;,\qquad \qquad a_n \;\sim \;
\left( \frac{l}{2} b\right)^n\, n!\;\;.
\label{c5}
\eeq
This (same-sign) factorial behavior manifests the presence of the IR
renormalon singularity \cite{thooft} which makes the series ill-defined at the
level of power corrections, $\sim (\Lam/Q)^l,\;\;\;l=1, 2,...\;$. On the other
hand, this is a deficiency of 
the purely perturbative expansion itself and has
nothing to do with the actual strong-interaction domain \cite{ir}; in
particular, the factorial growth persists even if the effective coupling never
becomes large but stays finite for all $k^2$, when the integral in \eq{c4} is
unambiguously defined \cite{ir,grun}.

It is important to note that the IR renormalons are 
absent in the Wilson OPE, when
the separation between the operators and the coefficient functions is done
according to the short-distance or long-distance origin of the contribution
\cite{fail}. They appear only at the attempts to define `purely perturbative'
and `purely nonperturbative' pieces of an observable. In this respect
processes with the heavy quarks do not have any peculiarity.

The IR renormalons in the HQET were addressed recently also in 
Refs.~\cite{ns,lms}. I
cannot agree with the message of \cite{lms} which claimed the 
IR renormalons
to be a peculiar feature of the {\rm effective} theories. 
On the contrary, the
IR renormalons are generally identical in QCD and its effective low-energy
limit, since their difference, by definition, lies only in the ultraviolet
domain. The structure of the IR singularities in perturbative QCD does 
not depend on whether
the QCD is a fundamental underlying theory, or is an effective low-energy
representation of a more general quantum system like GUT or a superstring
theory. Appearance of the IR renormalons in the perturbative calculations 
depends solely on the way how one
performs the OPE in a theory.

A certain peculiarity of the HQET in its standard formulation, in this respect,
is that it has an extra {\em spurious} $1/m_Q$ IR renormalon. It does not have
any deep reason but is merely rooted in the definition of the field variable 
routinely used in the HQET for the heavy quarks (for simplicity I ignore 
the spinor indices):
\beq
h_v(x)\;=\; {\rm e} \, ^{i \,(vx)\,m_Q^{\rm pole}}\cdot Q(x)
\label{c7}
\eeq
with the {\em pole} mass in the exponent. It is just this factor that brings in
the leading $1/m_Q$ renormalon singularity into the HQET perturbative series
for short-distance quantities. Since none of the relations between the
observables can depend on the phase definition used in the calculations, this
spurious effect always disappears in the relations among short-distance
observables \cite{pole}. At the technical level, however, it sometimes may look
not obvious.

In the Wilson OPE one defines an effective theory normalized at some scale
$\mu$; in particular, the basic parameter for the heavy quark physics is
$m_Q(\mu)$, which is the short-distance quantity, not sensitive to the physics
below $\mu$. The most convenient (but, certainly, not the only possible) choice
for the nonrelativistic field definition is then 
\beq
\tilde Q_{(\mu)}(x)\;=\; {\rm e} \, ^{i \,(vx)\, m_Q(\mu)}\cdot Q_{(\mu)}(x)
\label{c8}
\eeq
where the subscript $(\mu)$ denotes the normalization point. In this case one
has an operator equation of motion
\beq
i(vD)_{(\mu)}\tilde Q_{(\mu)}\;=\; 0\;+ \;{\cal O} \left(\frac{\mu}{m_Q} 
\right)\;.
\label{c9}
\eeq
In the perturbation theory the product of the operators in the l.h.s. has the
perturbative corrections $\sim \as^n\cdot \mu$ coming from both the operator
$D_\al = \partial_\al+g_s A_\al$ and from the interaction of $\tilde Q$ with
the gluon field. They combine to satisfy \eq{c9} to any order \cite{pole}. In
the standard HQET approach one insists \cite{neubpr,ns,lms} that no
perturbative correction is present from any separate source, and this unnatural
requirement leads to the IR renormalons in the matching with the QCD.

Based on the renormalon analysis, Neubert 
claimed \cite{update,ns} that the sum rule
(\ref{25})
cannot be correct, since $1/m_Q^2$ renormalons 
allegedly mismatch in it. Although 
in Wilson's OPE the IR renormalons are always
absent from any particular term, the IR renormalon calculus can still
be applied if the OPE relation is considered in the pure perturbation theory
itself, and
formally setting $\mu=0$ (which is technically possible to 
any finite order in
$\as$). This is nothing but a cross-check how the OPE works in the one-loop
perturbative calculations. Compared to Wilson's procedure, 
the limit $\mu\ra 0$, in particular, amounts to 
subtracting a `perturbative piece' from
the observable probabilities entering, for example, the phenomenological
side of the sum rules.
In this way the
perturbative terms obviously appear in the left-hand side of \eq{25} as well, 
and these
terms were ignored in \cite{update,ns}. 
Including them immediately balances -- not surprisingly -- the IR renormalons
in the both sides \cite{pert}.

On the other hand, the analysis of \re{ns} is instructive demonstrating once
more that HQET in that interpretation is not a closed theory once
nonperturbative effects are addressed. For example, the sum rules in their
literal form indeed cannot be derived in it.
Clearly, there is nothing wrong with the sum rules themselves, the reason is
merely non-existence (at the nonperturbative level) of HQET as an effective
quantum theory with $\mu$-independent matching coefficients, according to
prescriptions in its popular formulation \cite{neubpr}.

Leaving aside theoretical aspects of the IR renormalons, it is still
important to calculate the perturbative corrections in the practical 
applications of the 
OPE. The first such calculation for the sum rules was done in \cite{optical}
and then addresses again in a number of papers \cite{yakov,lig}. The ${\cal
O}(\as)$ terms appeared to be rather moderate. Based on the first terms 
$\sim b\as^2$ in the BLM series, it was suggested in 
\cite{lig} that the higher-order radiative
corrections to the sum rules are too large and allegedly make them next to
useless. These conclusions stemmed from
the alternative, non-Wilsonian use of the sum rules of \cite{vcb,optical} when 
the perturbative pieces as they literally emerge in the BLM calculation, are 
completely subtracted from all the entries. The large corrections then came
basically from integrating the running coupling in the domain
below the Landau singularity.
The numerical analysis in the OPE \cite{pert}, on the contrary, 
suggested a quite moderate impact of the radiative corrections.

According to \cite{lig}, the perturbative corrections to the sum rule of the
type of \eq{26} weaken the bound for the expectation 
value of the kinetic operator 
to such extent that it becomes non-informative. To appreciate the status of 
this statement,
one must realize that, in the quantum field theory,  the renormalized operators
can be defined in different non-equivalent ways; $-\lambda_1$ addressed in 
\cite{lig} is
known to be different from $\mu_\pi^2$. 
In general, the complete definition of an
operator at the nonperturbative level is a nontrivial task.
As a matter of fact, the only complete
definition of the kinetic operator $\bar{Q} (i\vec D\,)^2 Q$ given so far was
made in \cite{optical}. 
In
any case, an effective operator, intrinsically normalization-scale independent
as intended in HQET, cannot be defined in a QCD-like theory.
The inequality
$\mu_\pi^2>\mu_G^2$ always holds for the expectation value of the well-defined 
operator of \cite{optical}, for arbitrary normalization point. Moreover,
with this definition the perturbative normalization-point dependence has been
calculated there (see also \cite{pert}). 

As for $-\lambda_1$, a parameter in HQET, its definition 
beyond the classical level has
never been given; the procedure adopted in \cite{lig} reduces to an attempt
to completely subtract the `perturbative piece' of $\mu_\pi^2(\mu)$:
\beq
-\lambda_1= \mu_\pi^2(\mu)-c_1\frac{\as(\mu)}{\pi}\mu^2-
c_2\left(\frac{\as(\mu)}{\pi}\right)^2\mu^2 - ...
\label{34}
\eeq
(the method to calculate $c_i$ was elaborated in \cite{optical}). Yet 
such a program cannot be performed even in principle \cite{fail}.
The situation appears rather simple in the BLM
approximation, where all $c_i$ are readily calculated: the series, whose second
term was discussed in \cite{lig}, is divergent and sign-varying
\cite{pert}, so
using merely the second term is misleading for evaluation. 
Numerically, the perturbative subtraction looks as follows:
\beq
-\lambda_1= \mu_\pi^2(\mu)- 0.12\GeV^2 - 0.3\GeV^2 +...
\; \ge\; 0.4\GeV^2- 0.12\GeV^2 - 0.3\GeV^2 +\;...\; ,
\label{34a}
\eeq
with the ellipses denoting higher-order terms which already grow. Based on the
third term, it was concluded in \cite{lig} that no bound for the expectation
value of the kinetic operator can be derived. In what concerns the HQET
parameter $-\lambda_1$, it is perfectly true:
clearly there can be no definite bound established for a quantity that has 
not been -- and cannot be -- defined.

Concluding the discussion of the kinetic operator, I should emphasize
that the above subtleties are peculiar to the field-theory analysis. 
The inequality $\mu_\pi^2>\mu_G^2$ must hold in {\em any} 
QM model relying on a potential description not invoking additional 
degrees of freedom beyond the heavy quark and a spectator, if the
heavy quark Hamiltonian is consistent with QCD. Unfortunately, a failure to
realize this general fact is seen in a number of recent analyses.

A similar second-order BLM improvement 
of the sum rule (\ref{25}) for $F_{D^*}$
was also attempted in \cite{lig} and claimed to destroy the 
predictive power of
this relation (the first-order calculation had been performed in 
\cite{optical} as well). The calculation of the impact of the second-order
perturbative terms, however, was not done consistently, and the
actual effect is smaller \cite{pert}.

In reality, the perturbative corrections seem to be modest, at
least for reasonable values of the running coupling. To illustrate this
assertion without submerging into details, let me introduce the following 
notation:
\beq
\eta_A(\mu)\;\equiv \;\left[\xi_A(\mu)\right]^{1/2}\;\;.
\label{35}
\eeq
$\eta_A(\mu)$ can indeed be calculated in the perturbative expansion; for
example, to order $\as$ it is given by 
\beq
\eta_A(\mu)\;=\; 1+ 
\;\frac{\as}{\pi}\left[\frac{m_b+m_c}{m_b-m_c}\log{\frac{m_b}{m_c}} -
\frac{8}{3} + \frac{1}{3}\left(
\frac{\mu^2}{m_c^2} + \frac{\mu^2}{m_b^2} + \frac{2\mu^2}{3 m_c m_b}
\right) \right] \;.
\label{35a}
\eeq
The quantity $\eta_A(\mu)$ must be added to $\delta_{1/m^2}$ in the framework
of the OPE to calculate or bound $F_{D^*}$, instead of $\eta_A$ in the model
calculations. Then, at a reasonable choice
$\mu\simeq 0.5\GeV$, $\Lam=300\MeV$ (in the $V$-scheme) one has
\begin{eqnarray*}
\eta_A(\mu)\; = & 1  &  \;\;\;\mbox{ tree level}\\
\eta_A(\mu)\; = & 0.975 &   \;\;\;\mbox{ one loop}\\
\eta_A(\mu)\; = & 0.99  &  \;\;\;\mbox{ all-order BLM}
\end{eqnarray*}
Clearly, the effect of the calculated perturbative corrections is not
significant 
and $\eta_A(\mu)$ is very close to the value of $0.98$ adopted in the original 
analysis \cite{vcb}. The dependence of the BLM-resummed value of
$\eta_A(\mu)$ on $\mu$ and on the QCD coupling is 
illustrated in Fig.~3 ($\Lam$ is shown in the $\overline{\rm MS}$ scheme).

\begin{figure}
\vspace{5.9cm}
\includegraphics{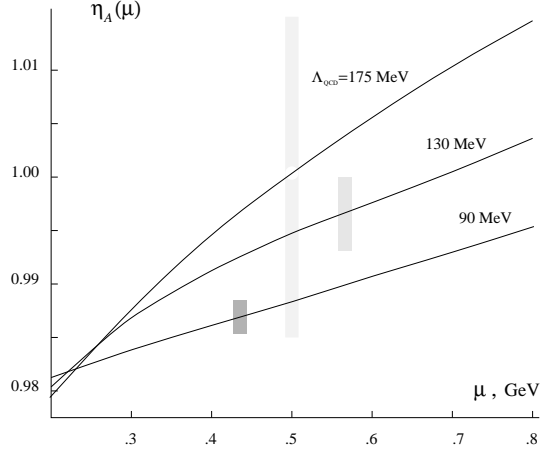}
\caption{The value of $\eta_A(\mu)$ for $\Lam^{\overline{\rm MS}} =90\MeV$,
$130\MeV$ and $175\MeV$. The value of $\eta_A(\mu)$ must
be used in the QCD-based calculations of the exclusive zero-recoil $B\ra D^*$
form factor when nonperturbative effects are addressed. The shaded bars show
the perturbative uncertainty irreducible without considering 
$1/m_c^3$ and $1/m_c^4$ effects; they give the lower bound for
the corresponding actual nonperturbative corrections.}
\end{figure}

Conceptually, the deficiency of the alternative usage of the sum 
rules of \res{vcb,optical} chosen in \cite{lig} is to gauge $\xi_A$ on
the value of $\eta_A^2$ as it has been defined in the HQET (the idea that
$\xi_A$ is to be identified with $\eta_A^2$ ascends to \cite{ns,update}): 
the value of $\eta_A$ is defined as an all-order `purely perturbative'
renormalization factor of the zero-recoil axial current. However, it is this
factor $\eta_A$ that is ill-defined, and only for this reason must the 
difference
between the stable Wilson coefficient $\xi_A$ and $\eta_A^2$ suffer from
large corrections. It is worth reiterating once more in this respect that, in 
reality, $\eta_A$
{\em cannot} be equal to a matching coefficient of $\bar c \gamma_\mu \gamma_5
b$ to a corresponding current in {\em any} effective field theory 
completely defined at
the level of the nonperturbative physics.

The accurate analysis thus shows that the  literal implementation of 
Wilson's OPE not only eliminates purely theoretical problems with the IR
renormalons,  but also makes the actual impact of the higher-order perturbative
corrections well-controlled.

\section{Bounds on the transition formfactors from analyticity and
unitarity}

The kinematic point of zero recoil in the $b\ra c$ semileptonic
transitions is the least uncertain one 
for the theoretical analysis. On the other
hand, experimentally the statistics is limited here. Therefore, information
about the $q^2$-dependence of the formfactors is very important. However, 
is has a dynamical origin and is unknown {\em a priori}. Indeed, if the $B$
meson is relatively `soft', i.e. the typical velocity of the effective light
degrees of freedom is small -- as would be the case in the nonrelativistic
quark models -- then the formfactor (the IW function) changes at the velocity
transfer much less than unity, and this implies large derivatives 
${\rm d}F(\vec v^{\,2})/{\rm d}\vec v^{\,2}$, ${\rm d}^2F(\vec v^{\,2})/({\rm
d}\vec v^{\,2})^2$, etc. If, on the contrary, the bound state is relatively
simple and the light degrees of freedom are ultrarelativistic, these 
derivatives are expected to be 
of order unity. Can something be stated about them without 
actual information about the bound state dynamics? I will argue below that the
answer is negative.

An idea was put forward a few years ago to constraint the IW function using,
mainly, general analyticity and unitarity properties of the formfactors in the
heavy quark limit \cite{taron}. They turned out 
to be non-informative due to the
presence of the $Q\bar Q$-bound states below the threshold of the open flavor
production. This approach was resurrected recently in a number of publications
\cite{lebed,caprini}. In particular, 
it was claimed in \cite{caprini} to have established an intriguing relation
between the slope and the curvature of the formfactor near the zero recoil
kinematics. However, there are essential flaws in that analysis.

The general idea is to consider the transition formfactor for some $\bar c\, b$
current between $B$ and $D$ mesons as an analytic function of the momentum
transfer $q^2$:
\beq
\matel{D(p+q)}{\:\bar c\,b\,(0)\:}{B(p)}\;=\; 2\sqrt{M_B M_D} \; F_S(q^2) 
\label{n3}
\eeq
where, for simplicity, we can consider the scalar current not yielding
irrelevant Lorentz indices.

The function $F_S$ in the physical $t$-channel 
domain $q^2 \le t_{\rm zr}= (M_B-M_D)^2$
describes the decay amplitude and is real. On the other hand, analytically
continued to large $q^2 > (M_B+M_D)^2$ it describes the exclusive production
of $B\bar D$ in ${\rm e^+e^-}$-annihilation--type process and thus, in
principle, is also an observable quantity. $F_S$ is complex there. Under the 
usual
assumptions of analyticity one can write the dispersion integral for $F_S$:
$$
F_S(q^2)\;=\; \frac{1}{\pi}\,\int\; 
\frac{\Im F_S(t+i\epsilon)}{t-q^2-i\epsilon}\,
dt\;=\; 
\frac{1}{\pi}\,\int_{t<(M_B+M_D)^2}\; \frac{\Im F_S(t)}{t-q^2-i\epsilon}\,dt\;
+ 
$$
\beq
+\;
\frac{1}{\pi}\,\int_{t>(M_B+M_D)^2}\; \frac{\Im
F_S(t)}{t-q^2-i\epsilon}\,dt\;\equiv \; F_S^{(r)}(t)\;+ F_S^{(c)}(t)\;.
\label{n4}
\eeq

If the sub-threshold contribution from $\Im F_S(t)$ for $t<(M_B+M_D)^2$ were
small, the formfactor would be saturated by the second integral, $F_S\simeq
F_S^{(c)}$. In that case the formfactor would be essentially constrained: 
since the
exclusive production of $B\bar D$ must not yield the production cross section
in this channel larger than the total hadronic one which is bound 
theoretically, $F$ cannot be too 
large in the wide domain. If one additionally has the normalization 
\beq
F_S(t_{\rm zr})\;=\;1
\label{n6}
\eeq
(which holds at large $m_{b,c}$) 
then only little room would have been 
left and, for example, an almost functional 
relation between $F^\prime_S (t_{\rm zr})$ and $F^{\prime\prime}_S(t_{\rm zr})$ 
emerged \cite{caprini}.

This reasoning, however, crucially relies on the insignificance of the
sub-threshold support of $\Im F(q^2)$; on the other hand, it is generally known
not to hold for heavy quarks. On the contrary, the common wisdom is that the
first resonances dominate the formfactor. The idea of \re{caprini} was to
consider the formfactor $F_S$ which does not have a contribution from the 
well-studied vector quarkonia. It was assumed that no clear-cut resonances 
below the $B\bar D$ threshold exist in the corresponding channel.

This basic assumption is erroneous, however. The sub-threshold scalar $0^{++}$ 
states are present both in the $\bar c c $ and $\bar b b$ systems ($\chi_{c,b}$
states) and, according to the general theorems, must be also in the
$\bar b c$ channel.\footnote{In any case it is ensured by the heavy flavor
symmetry which is inherent in the analysis of \cite{caprini}.} They are
extremely narrow since lie below the 
open-flavor thresholds, and are expected to dominate the 
formfactor.\footnote{It can be argued that the
contribution of the continuum domain $q^2>(M_B+M_D)^2$ itself falls short of
$1$  for $F_S$ at the zero recoil point $q^2=(M_B-M_D)^2$.}

To account for the possible effect of the domain $q^2<(M_B+M_D)^2$, 
\re{caprini} equated the whole sub-threshold contribution with the possible
nonresonant one and 
considered a smooth model for $\Im F_S$ below the threshold:
\beq
\Im F_S(t)\;\simeq \; \frac{M_B+ M_D}{2\sqrt{M_B M_D}}\;
\frac{C}{\sqrt{t_0}}\, \sqrt{t-(M_{B_c}+m_\eta)^2}\;\;,
\;\;\;\;(M_{B_c}+m_\eta)^2 <t<(M_B+M_D)^2
\label{n7}
\eeq
with $C$ of order 1. As explained above, such a functional form is irrelevant
below the threshold. Moreover,
it is easy to calculate its contribution to the zero-recoil 
value $F_S((M_B\!-\!M_D)^2)\,$:
\beq
F_S^{(r)}(0)\;\simeq\;  0.008\, C
\label{n8}
\eeq
(recall that in the usual pole-dominance models this value must be just close 
to unity). Thus, the effect of the states lying below the open flavor threshold
was underestimated by two orders of magnitude.

The analysis of \re{caprini} applied to the scalar $\bar{c}\,b$ formfactor.
Obtaining model-independent information about any formfactor would be 
very interesting by
itself from the theoretical viewpoint. Phenomenologically,
however, one needs to know the axial-vector formfactors, or those in the heavy
quark limit $m_Q \ra \infty$. In the heavy quark limit all these formfactors
are related to the IW function; unfortunately, the $1/m_c$ corrections in the
derivatives are too significant. Although they were claimed to be very mild, in
reality it was based only on some perturbative estimates.
On the other hand, the main,
nonperturbative $1/m_c$ corrections are known to be large \cite{vain},
$\sim 50\%$.\footnote{Although it was admitted in \cite{caprini} that the
impact of the $1/m_c$ corrections was missed in the analysis, this
qualification was somehow absent from the phenomenological claims.}

Can these problems be cured? One could have tried to consider similar
dispersion
relations at arbitrary masses $m_{c,b}$. Decreasing $m_c$ would soften the
problem of the sub-threshold physics -- however, the corrections in the
short-distance expansions and to the necessary heavy quark symmetry relations
are already marginal. 
Attempts to increase $m_c$ to tackle the $1/m_c$
corrections magnify the role of the sub-threshold behavior of the
formfactor.

It is important that, contrary to the forward scattering amplitudes or current
correlators, the phase of the formfactor has only an indirect physical meaning,
and below the threshold is a very unphysical quantity not observable directly.
For this reason its behavior in this domain can be rather odd and crucially
depends on irrelevant details of the strong interactions which we know
almost nothing about; by no means, for example, $\Im F$ must be positive for
heavy quarks.

A very instructive analysis had been given a few years ago in the inspiring
papers  \cite{jaffe} which, unfortunately, are to a large extent forgotten  
in the recent literature.
It was argued there that, just if the heavy
flavor transition formfactor is governed by relatively soft clouds of light
degrees of freedom, in the $s$-channel one should expect 
a complicated oscillating behavior of the production formfactor near the 
threshold; it can
be simple if the transition formfactor is a slowly varying function (a
`hard-core' bound state) and then the lowest-pole dominance is expected to 
work.

One has to conclude, once again, that without actual dynamical input, the 
general
analyticity and unitarity relations cannot {\em per se} yield useful 
information about 
the transition formfactors of interest. From the phenomenological perspective,
the analysis \cite{caprini} grossly underestimated both the sub-threshold 
contributions and the corrections \cite{vain} to the heavy quark
symmetry constraints it relied on. 
The relation between the slope and the curvature claimed
there rather should not be used in experimental analyses for deriving
model-independent results.

The possibility to employ additional dynamical information, e.g. in the form
of the lattice calculations of the formfactors at some intermediate points
\cite{lelush}, must be explored in order to revive 
this approach.

\section{The semileptonic branching ratio}

In the QCD-based heavy quark expansion the total nonleptonic decay widths of
heavy flavors can be treated in the same way as semileptonic or radiative ones;
the difference appears only at a quantitative level. The
statement about the absence of $1/m_Q$ corrections applies as well, and the
leading corrections $\sim 1/m_Q^2$ have been calculated \cite{buv,bs,bbsuv}.
The overall semileptonic branching ratio ${\rm BR_{sl}}(B)$ seems to be of a
particular practical interest: while the simple-minded parton estimates yield
${\rm BR_{sl}}(B)\simeq 15\%$ \cite{parton}, experiment gives smaller
values, ${\rm BR_{sl}}(B)\simeq 10.5$--$11.5\%$ (the 
current situation is reviewed
by T.~Browder).

The generic scale of the $1/m_b^2$ effects in the widths of individual 
nonleptonic channels is about
$15\%$; however, the particular $V$--$A$ chiral and color structure of weak
currents significantly suppresses it; moreover, an 
additional
cancellation between different parton channels occurs, and one literally gets
\cite{buffling}
\beq
-\Delta_{1/m^2}\left({\rm BR_{sl}}\right)\; \lsim \; 0.5\%\;\;.
\label{40}
\eeq
Estimated $1/m_b^3$ corrections do not produce a significant effect either 
\cite{mirage}. As a result, most of the attention was paid to a more accurate
treatment of the perturbative corrections, in particular, accounting for the
effect of the charm mass in the final state. It was found
\cite{bagan,volbr} that the nonleptonic width is indeed boosted up.

Since the inclusive widths are expanded in inverse powers of 
energy release, one {\em a priori} expects larger corrections or even a
breakdown of the expansion and violation of duality in the channel $b\ra c\bar 
c s(d)$; however, this channel can be isolated \cite{buv} via charm counting, 
i.e. measuring the average number of charm quarks per beauty decay. The
original 
experimental estimate $n_c \lsim 1.15$ did not allow one 
to attribute the apparent
discrepancy to this channel, and gave rise to the so-called `${\rm
BR_{sl}}$ versus $n_c$' problem.

The perturbative corrections in the $b\ra c\bar u d$ itself cannot naturally 
drive $\rm BR_{sl}$ below $12.5\%$; the calculation 
for the $b\ra c\bar c s(d)$
channel is less certain and, in principle, admits increasing the width by a
factor of $1.5$ or even $2$, leading to $n_c\simeq 1.25$--$1.3$. In the latter
case a value of $\rm BR_{sl}$ as low as $11.5\%$ can be accommodated.

The experimental situation with $n_c$ does not seem to be quite settled yet:
$$
n_c =1.134\pm 0.043 \qquad \mbox{(CLEO) }\;,\qquad \qquad 
n_c =1.23\pm 0.07\qquad \mbox{(ALEPH) }\;.
$$
In a recent analysis \cite{dunietz} it was argued that consistency requires a
major portion of the final states in the $b\ra c\bar c s$ to appear as 
modes with the wrong-sign $D$'s and kaons but $D_s$, which previously 
escaped proper attention,  
and this allowed for a larger value $n_c\simeq 1.3$ needed to
resolve the problem with $\rm BR_{sl}$. The dedicated theoretical 
consideration  
\cite{bsu} shows that, indeed, the dominance of such modes is natural and this
possibility does not require a significant 
violation of duality. 
Thus, if
a larger value of $n_c\simeq 1.25$ is confirmed experimentally,
the problem of $\rm BR_{sl}$
will not remain.

Even within this successful scenario a concern can be raised if a strong
QCD-enhancement of a tree-level--unsuppressed decay mode, possibly achieved in
the one-loop calculations, is trustworthy: one would need to boost the
tree-level $b\ra c\bar c s$ width by a factor of two. One should note,
however, that theoretically the situation maybe not so pessimistic: the tree
level estimate of ${\rm BR_{sl}}(b\ra c\bar c s) \simeq 15\%$ \cite{parton} 
refers 
only to using the large values of the quark masses close to their pole
values. Using the short-distance masses like $m_b\simeq 4.5\GeV$ always yielded 
a larger starting value of the branching for this channel. 
Without addressing the strong interaction effects it must be viewed,
conservatively, as an uncertainty of the simplest partonic predictions.
However, it is more consistent, from the viewpoint of the OPE, 
to always use the 
values of the parameters entering at a particular level of computations; in
the case of the $B$ widths it means that the smaller short-distance masses are 
preferable as the starting approximation. If one accepts this attitude, 
the actual corrections to the decay rate found at the one-loop level are
rather mild and do not prompt immediate concerns about the theoretical control 
over the perturbative expansion. The observed large corrections to ${\rm
BR_{sl}}(b\ra c\bar c s)$ are then merely a response to an 
improper zeroth-order approximation.

\section{Lifetimes of beauty particles}

The QCD-based HQE provides a systematic framework for calculating the total
widths of heavy flavors, which are not amenable to the traditional methods of
HQET. The only assumption is that the mass of decaying quark 
(actually, the energy release) is sufficiently large; 
for a review, see \cite{stone}.

A thoughtful application of this expansion to charm particles demonstrated
that the actual expansion parameter appeared to be too low to ensure a 
trustworthy accurate description, so that {\em a priori} one expects only 
emergence of the 
qualitative features. Surprisingly, in many cases the expansion
works well enough even numerically. In particular, it predicts $\tau_{D^+}
\approx  2\tau_{D^0}$ as an effect of Pauli interference in $D^+$ and
suggests $\tau_{D_s} \approx  \tau_{D^0}$ -- simultaneously correctly  
predicting the semileptonic fraction in $D$ mesons. A similar
agreement seems to be observed in charm baryons (for a recent review, see
\cite{bigdor}).

Applying the expansion to beauty particles one expects a decent
numerical accuracy, although the overall scale of the preasymptotic effects is
predicted to be small and makes a challenge to experiment:
\beq
\begin{array}{cclcl}
\tau_{B^-}/\tau_{B^0} & \simeq & 1+0.04
\left(\frac{f_{B}}{180\,\rm MeV}\right)^2 &\cite{mirage} &
\qquad \mbox{ EXP: } \qquad 1.04\pm 0.04\\

\overline \tau_{B_s}/\tau_{B^0} & \simeq & 1+{\cal O}(1\%) &
\cite{mirage} & \qquad \mbox{ EXP: } \qquad 0.97\pm 0.05\\

(\tau_{B_s^L}-\tau_{B_s^S})/\tau_{B_s} & \simeq & 0.18 \left(
\frac{f_{B_s}}{200\,\rm MeV}\right)^2 & \cite{vsku} & \\

\tau_{\Lambda_b}/\tau_{B^0} & \approx &  0.9 & &
\qquad \mbox{ EXP: } \qquad 0.78\pm 0.06 \\
\end{array}
\label{lam}       
\eeq
These differences appear mainly as $1/m_b^3$ corrections and, depending on 
certain
four-fermion matrix elements, cannot be predicted at present very accurately.
In particular, it refers to the preasymptotic effects in baryons. For mesons
the estimates are based on the vacuum saturation approximation, which cannot be
exact either. The 
impact of non-factorizable terms has been studied a few years ago
in \cite{WADs} and possibilities to directly measure the matrix elements in
future experiments were suggested.

The apparent agreement with experiment is obscured by the 
reported lower values of
$\tau_{\Lambda_b}$. Since the baryonic matrix elements
are rather uncertain, quite a few model estimates have been done 
(see the recent one \cite{Rosner}, and references therein). All 
seem to fall short; 
however, this might be attributed to 
deficiencies of the simple quark model. It was
shown \cite{boost} using quite general arguments that, irrespective of the
details, one cannot have an effect exceeding $10-12\%$ while residing in the
domain of validity of the standard $1/m_Q$ expansion itself; the natural
`maximal' effects that can be accommodated are $\sim 7\%$ and $\sim 3\%$
for weak scattering and interference,  respectively.

Thus, if the low experimental value of $\tau_{\Lambda_b}$ 
is confirmed, it will require a certain
revision of the standard picture of the heavy hadrons and of convergence of the 
$1/m_Q$ expansion for nonleptonic widths.

\section{$1/m_Q$ expansion and duality violation}

The problem of duality violation attracts more and more attention of those who
study the heavy quark theory; a recent extensive discussion was given in
\cite{inst}. The expansion in $1/m_Q$ is asymptotic. There are basically two
questions one can ask here: what is the onset of 
duality, i.e. {\em when} does the
expansion start to work? The most straightforward approach was undertaken in
\re{bm}, and no apparent indication toward an increased energy scale was
found. Another question, {\em how} is the equality of the QCD
parton-based predictions with the actual decay rates achieved, was barely
addressed. Though a relevant example of such a problem is easy to
give.

The OPE in full QCD ensures that no terms $\sim 1/m_Q$ can be in the widths 
and the
corrections start with $1/m_Q^2$. However, the OPE {\em per se} cannot 
forbid a
scenario where, for instance,
\beq
\frac{\delta \Gamma_{H_Q}}{\Gamma_{H_Q}} \;\sim\ \;C \;
\frac{\sin{(m_Q \rho)}}{m_Q\rho}
\;,\;\;\; \rho \sim \Lam^{-1}
\;.
\label{60}
\eeq
In the actual strong
interaction, $m_b$ and $m_c$ are fixed and not free parameters, so, 
from the practical viewpoint these types of corrections are 
not too different -- but the difference is profound in the theoretical 
description! It reflects specifics of the OPE in Minkowski space, and such
effects can hardly be addressed,
for example, in the lattice simulations. Their control requires 
a deeper
understanding of the underlying QCD dynamics beyond the knowledge of first few 
nonperturbative condensates.

In fact, the literal corrections of the type of \eq{60} are hardly possible; the
power of $1/m_Q$ in realistic scenarios is larger, and these duality 
corrections must be eventually exponentially suppressed though, probably, 
starting at a higher 
scale \cite{inst}. But a theory of such effects is still in its embryonic
stage and needs an additional experimental input as well.

\section{Conclusions}

I would like to conclude this 
admittedly incomplete review of the theory and applications
of the heavy quark expansion by the following general remark. It was sometimes
said already a few years ago that the theory of heavy quark decays was
basically completed and only a few technical refinements remained to be done.
The subsequent development clearly showed that the potential for
further studies was vast. Regarding the current status, I believe that, although
the basic principles of the HQE in QCD are formulated and well studied, we
still have many interesting things to understand and calculate even in the
original, the most standard fields of applications -- before one can actually
consider the theory completed.

\vspace*{0.5cm}

{\bf Acknowledgements:}\hspace*{.5em} I am grateful to the organizers of 
CRAD `96 for the opportunity to participate in this
Symposium.
The financial support and hospitality of the CERN Theory Division are 
gratefully 
acknowledged. 
The collaboration with I.~Bigi, M.~Shifman and A.~Vainshtein was highly
beneficial; I am also thankful to M.~Voloshin for many clarifications.
This work was supported in part by NSF under the grant
number PHY 92-13313.

\end{document}